# Tropical Cyclone Forecasting Applications of the GOES WMSI

Kenneth L. Pryor
Center for Satellite Applications and Research (NOAA/NESDIS)
Camp Springs, MD

## 1. Introduction

The Geostationary Operational Environmental Satellite (GOES) sounder-derived Wet Microburst Severity Index (WMSI) (Pryor and Ellrod 2004) was originally developed and implemented to assess the potential magnitude of convective downbursts over the central and eastern continental United States. The WMSI algorithm incorporates convective available potential energy (CAPE), to parameterize static instability, as well as the vertical theta-e (equivalent potential temperature) difference (TeD) between the surface and mid-troposphere to infer the presence of a mid-level dry air layer. It has been noted with two recent landfalling hurricanes over the Florida Gulf of Mexico coast that the GOES WMSI product accurately predicted downburst magnitude associated with convective bands and the remnant eye walls. A tropical cyclone can be considered to be a circular mesoscale convective system, comprised of clusters of convective storms. As a convective system, it is expected that tropical cyclones can profilically produce downbursts. In fact several severe downbursts were observed in association with the pre-hurricane squall line and remnant eye wall of Hurricane Charley (August 2004) and the remnant eye wall of Hurricane Wilma (October 2005). This paper will discuss the evolution of downburst production associated with the two hurricanes as well as the effectiveness of the GOES WMSI product in predicting downburst potential.

## 2. Methodology

GOES WMSI product imagery was collected over a 24 hour period prior to the landfall of each tropical cyclone and validated against conventional surface data as recorded in METAR observations. Observed surface wind gusts associated with downbursts produced by the convective bands and remnant eye walls of each hurricane were compared to the most recent representative WMSI values for each event, assuming no change in environmental static stability and air mass characteristics had taken place. Next Generation Radar (NEXRAD) base reflectivity imagery (level II) from National Climatic Data Center (NCDC) was utilized to verify that observed wind gusts were convective in nature. Particular radar reflectivity signatures, such as the bow echo and the rear-inflow notch (RIN) (Przybylinski 1995), were effective indicators of the occurrence of downbursts.

## 3. Downburst Generation

a. Hurricane Charley

During the afternoon of 13 August 2004, at approximately 2032 UTC, intense Category 3 Hurricane Charley made landfall over the west coast of Florida between Port Charlotte and Punta Gorda. Hurricane Charley was one of the strongest tropical systems to affect southwestern Florida in over a decade. However, the effects of Charley were experienced over 12 hours earlier in south Florida as a pre-hurricane squall line tracked northward over the region, producing strong surface winds, especially over the greater Miami-Fort Lauderdale metropolitan area. A pre-hurricane squall line typically develops 200 to 500 miles ahead of the storm center and can be characterized by intense convection that generates strong downbursts. Measured wind reports associated with the squall line are indicated in the Table. Predicted wind gust speeds are based on linear regression as presented in Pryor and Ellrod (2004).

| Table. Measured Wind Speed vs. GOES WMSI | | | | |
|---|---|---|---|---|
| Time (UTC) | Location | Measured(kt) | WMSI | Wind Gust Potential (kt) |
| 0542 | Naples (KAPF) | 42 | 109 | 50-64 |
| 0545 | Pompano Beach (KPMP) | 49 | 141 | 50-64 |

A pre-hurricane squall line developed over the Florida Straits during the evening of 12 August prior to Hurricane Charley's landfall over western Cuba. GOES WMSI effectively predicted the relative strength of convective downbursts associated with this tropical system. GOES WMSI imagery observed the motion of the squall line from the Florida Straits northward into south Florida, finally weakening over central Florida by 1000 UTC.

The WMSI image at 0415 UTC 13 August 2004, displayed in Figure 1, revealed WMSI values in excess of 100 over southern Florida, ahead of the pre-hurricane squall line. The air mass in place over south Florida was potentially unstable, as portrayed by the high WMSI values into which the squall line was propagating. The strong instability that resulted in intense convection was signified by the appearance of overshooting tops in the enhanced infrared imagery. Also apparent was the presence of a mid-tropospheric layer of dry (low theta-e) air that could be entrained into the downdraft of mature convective cells and result in subsequent downdraft acceleration and downburst development. Thus, large WMSI implied potential for the development of strong updrafts and heavy precipitation and the subsequent development of intense convective downdrafts and downbursts. Accordingly, a downburst wind gust of near 50 knots at Pompano Beach was observed 90 minutes later (0545 UTC).

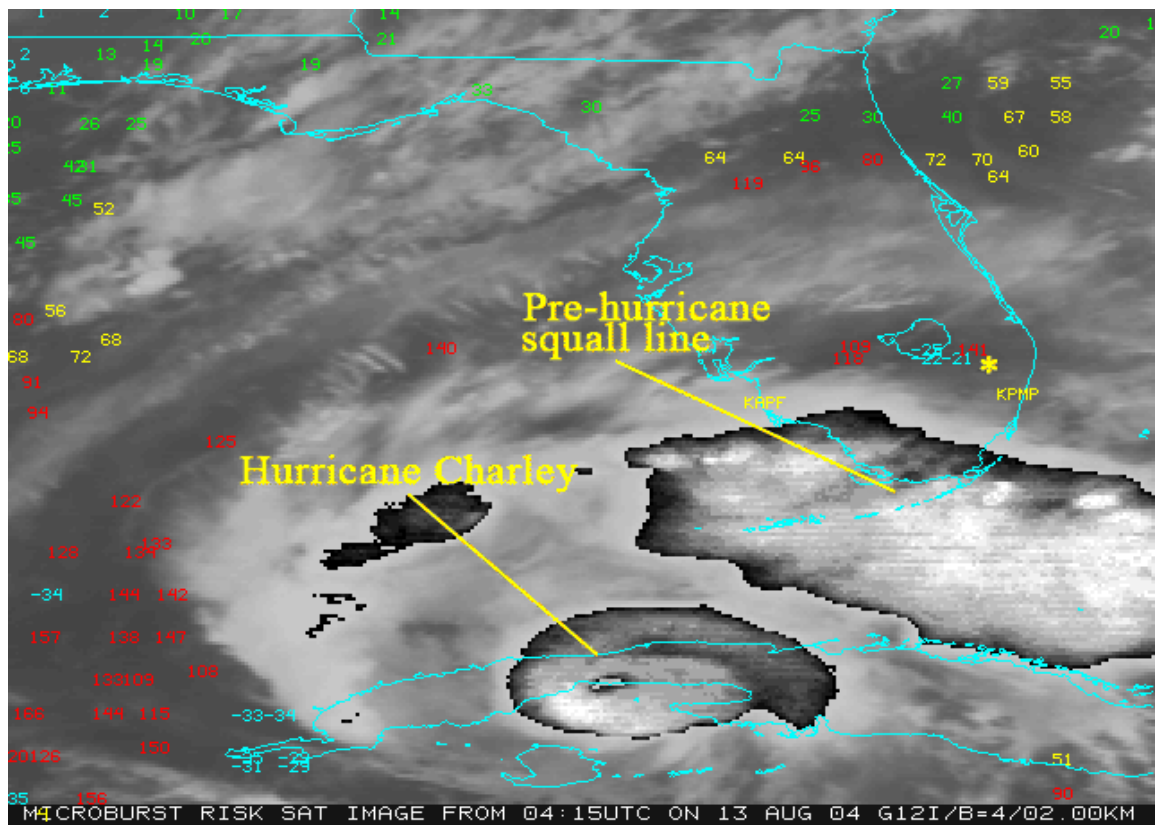

Figure 1. GOES WMSI image at 0415 UTC 13 August 2004.

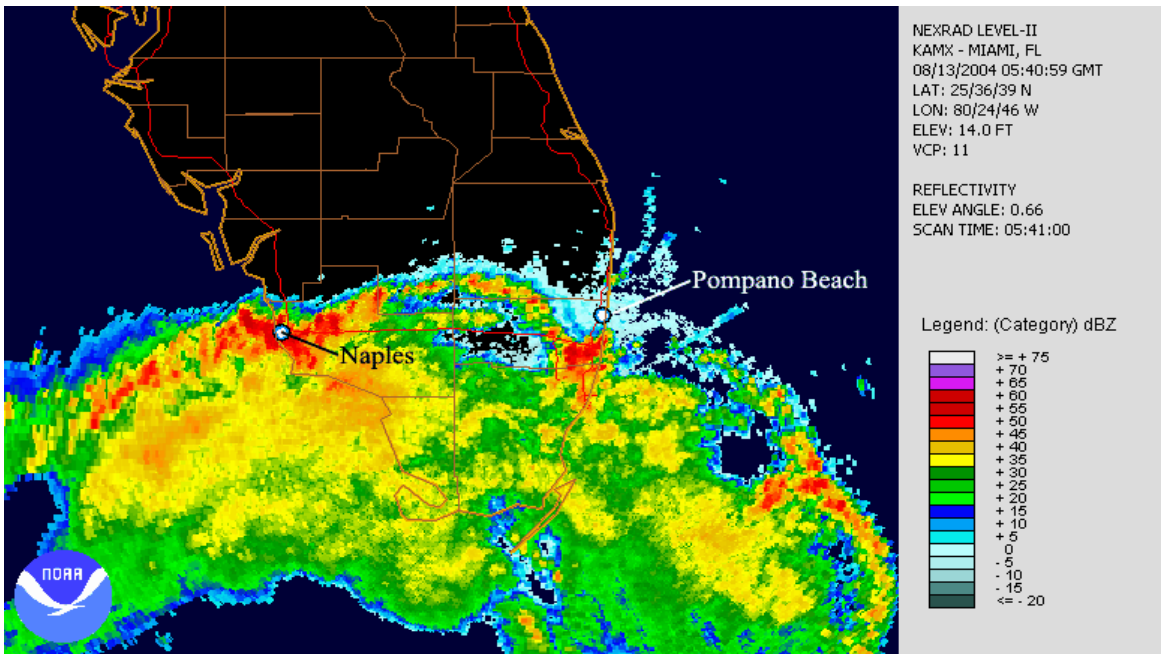

Figure 2. Miami, Florida NEXRAD reflectivity imagery at 0540 UTC 13 August 2004.

Miami, Florida NEXRAD (KAMX) reflectivity imagery clearly displayed in Figure 2 distinctive bow echoes (Weisman 1993; Przybylinski 1995) along the convective line that were associated with the downbursts that occurred in Naples and Pompano Beach. Intense low-level reflectivity gradients were evident along the leading edge of the bowing segments of the convective line, where low-level convergence was maximized. Also displayed were rear-inflow notches (RINs) along the convective line, suggesting that rear-inflow jets (RIJs) were penetrating into the convective cells' trailing edge (Przybylinski 1995). The RIJs were feeding low theta-e air from the rear of embedded convection cells and were thus instrumental in the generation of intense downdrafts that were observed at the surface as downbursts.

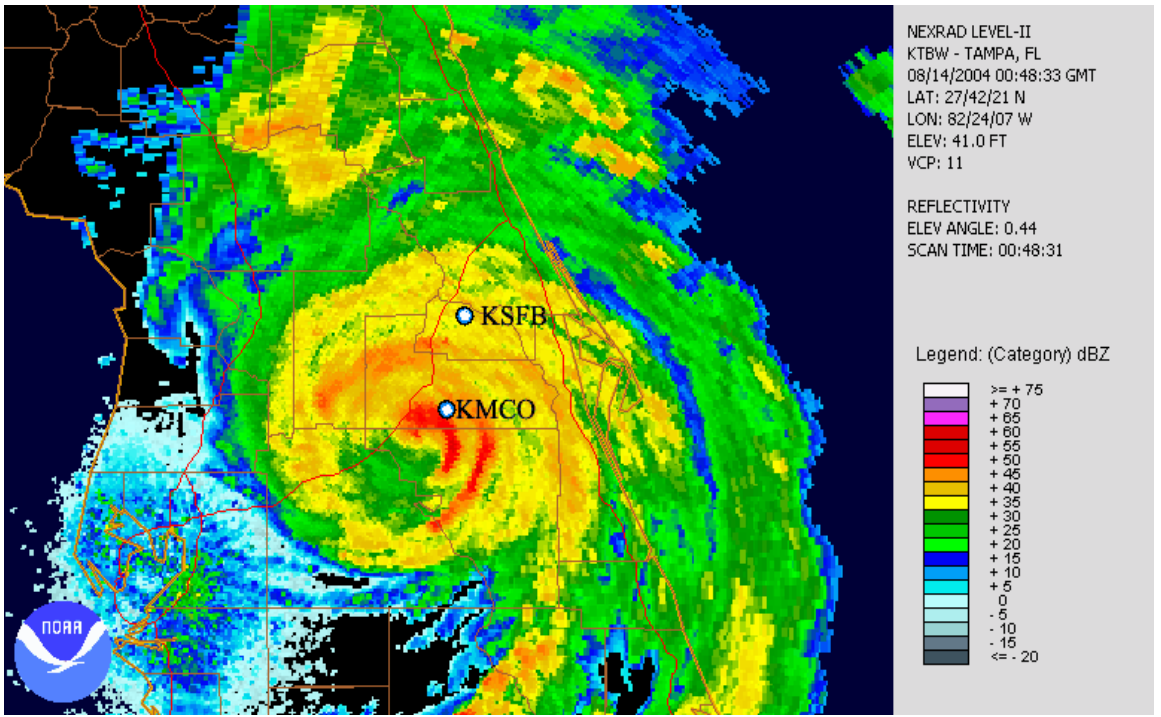

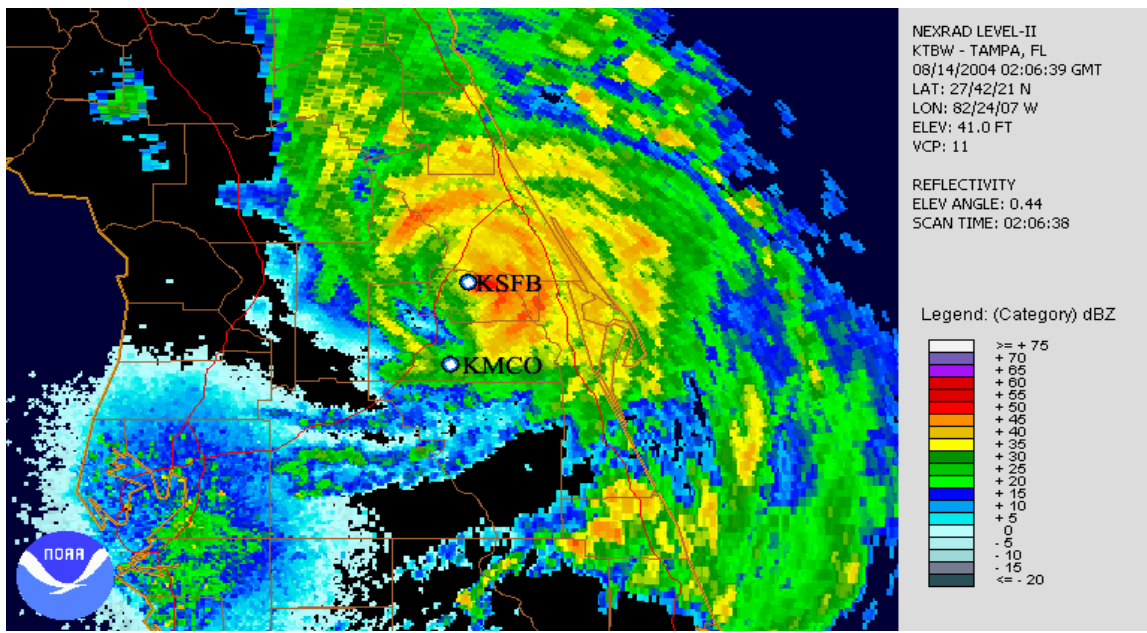

Figure 3. Tampa, Florida NEXRAD reflectivity imagery at 0048 UTC (top) and 0206 UTC (bottom) 14 August 2004. Markers indicate location of Orlando International Airport (KMCO) and Sanford (KSFB).

A distinct eye wall signature was apparent with Hurricane Charley well inland until approximately 2330 UTC. After 0000 UTC, as displayed in Figure 3, the eye wall rapidly evolved into a quasi-bow echo configuration, with a distinct northern bookend vortex present as the bow echo moved over the Orlando metropolitan area between 0048 and 0222 UTC. A peak wind gust of 80 knots was observed as the bow echo and associated bookend vortex passed over Sanford at 0209 UTC. Interestingly, the development and evolution of the bow echo from the remnant eye wall occurred in a region of high WMSI values, as indicated in Figure 4. A well-defined peak in wind speed, as apparent in a meteogram in Figure 7 from Sanford confirmed that the observed wind gust was in fact the result of an intense convective downburst. Atkins and Wakimoto (1991) connected this sharp wind speed peak within convective storms to the occurrence of wet microbursts or downbursts associated with heavy precipitation and high radar reflectivities.

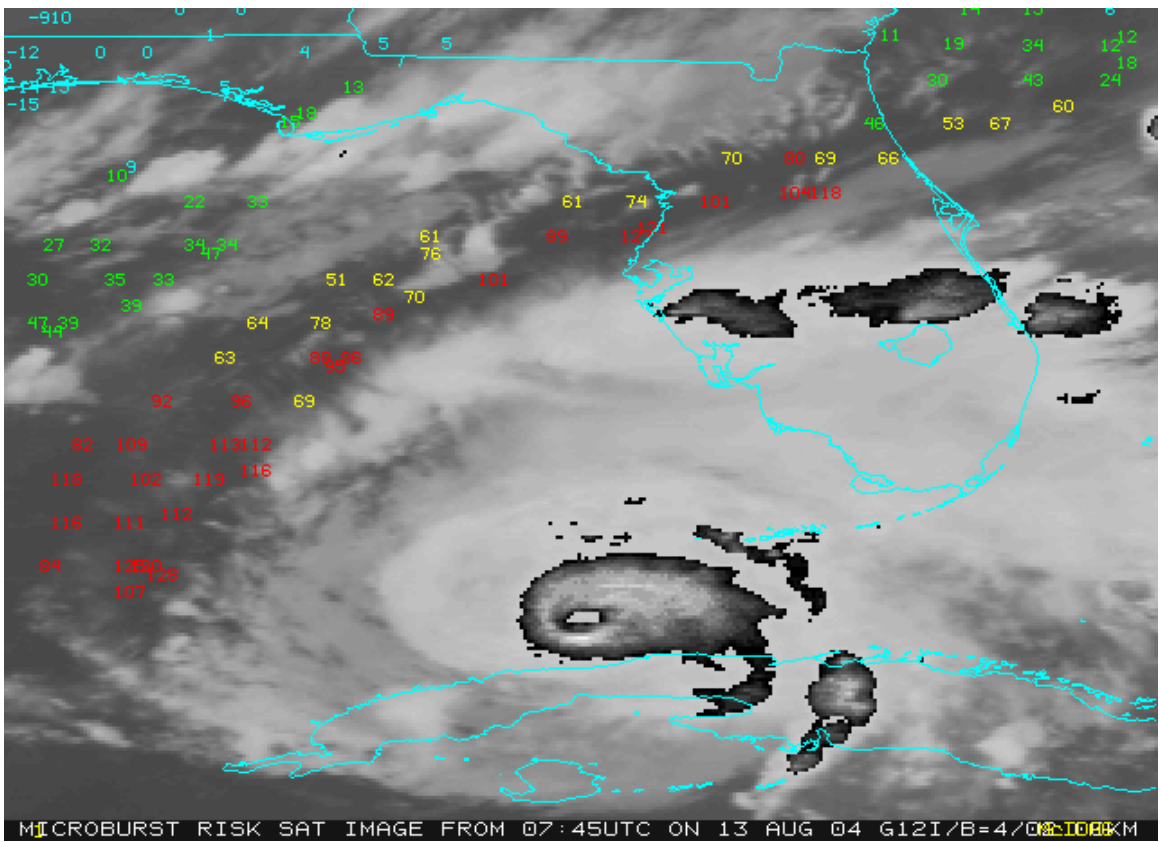

Figure 4. GOES WMSI image at 0745 UTC 13 August 2004.

b. Hurricane Wilma

During the morning of 24 October 2005, between 1000 and 1100 UTC, landfall of category two Hurricane Wilma occurred over the Gulf of Mexico coast of southwestern Florida near Marco Island followed by a northeast track to Jupiter Inlet. The Geostationary Operational Environmental Satellite (GOES) Wet Microburst Severity Index products of the previous afternoon, shown in Figure 5, indicated high WMSI values, corresponding to convective wind gust potential of 50 to 64 knots, over southwestern Florida, near the location of landfall. In addition, widespread convective downburst activity associated with the hurricane's eye wall was observed around the time of landfall. In fact, a downburst wind gust of 65 knots, associated with the eye wall, was observed at Naples Airport, near the location of the high WMSI value indicated in Figure 5.

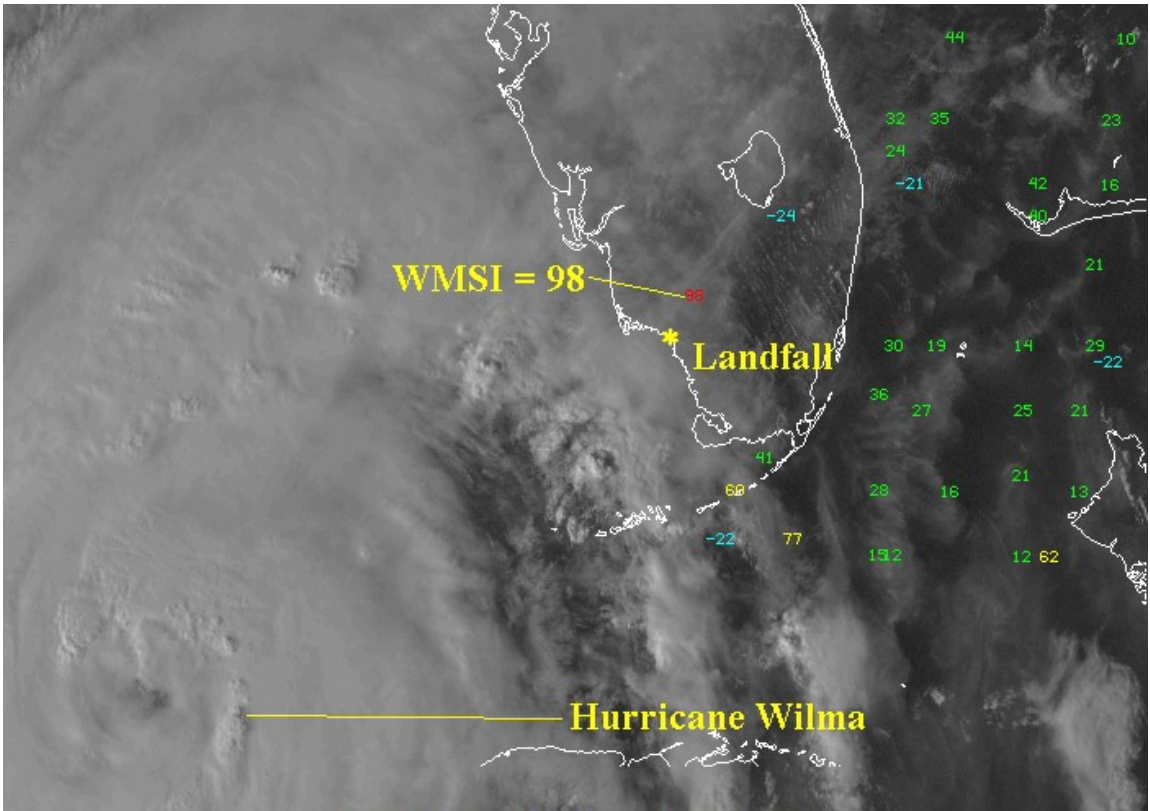

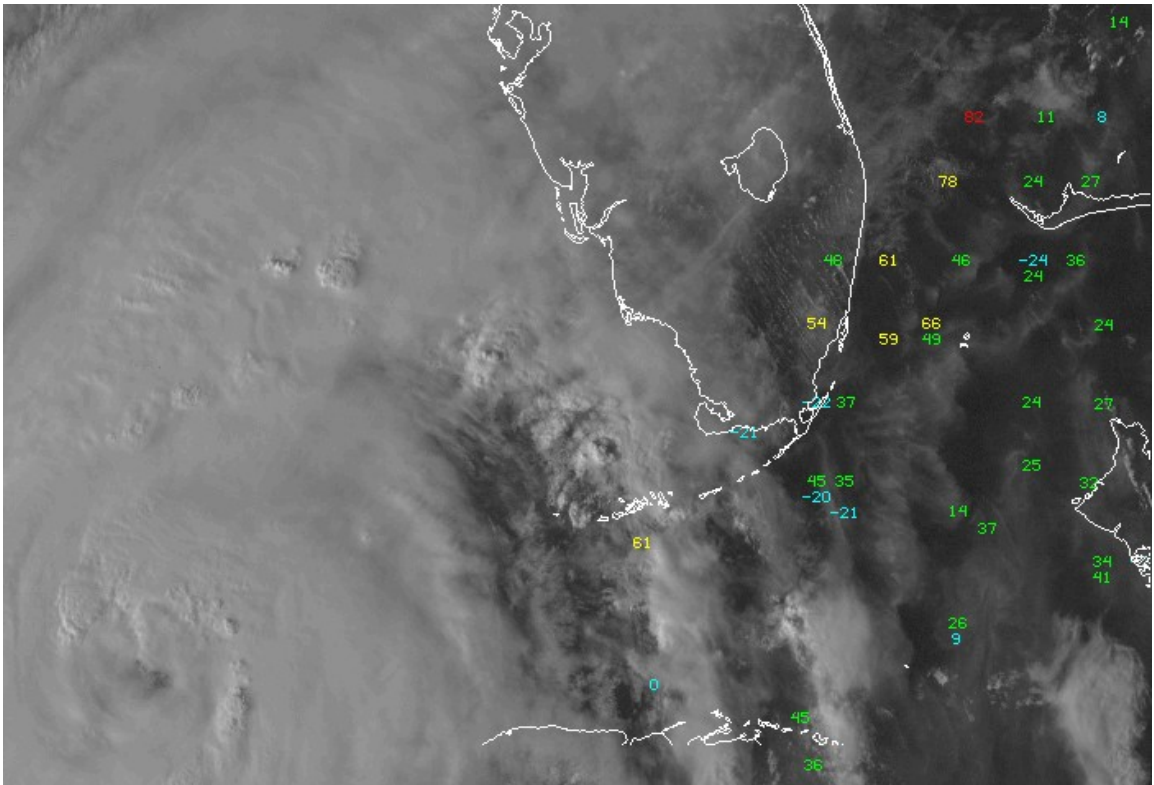

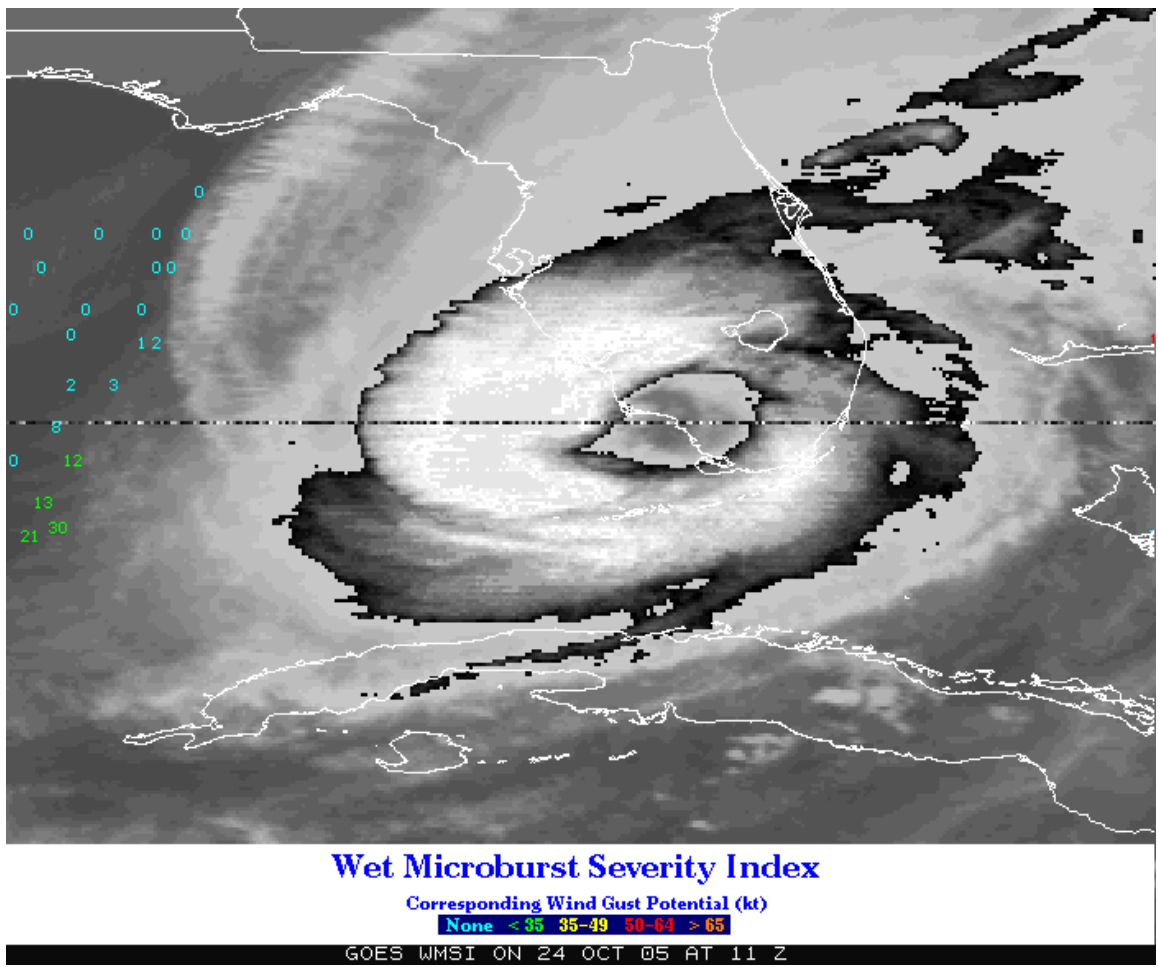

Figure 5. GOES WMSI images at 1900 UTC (top), 2100 UTC (center) 23 October 2005, and at the time of landfall, 1100 UTC 24 October 2005 (bottom).

Miami NEXRAD imagery (not shown) displayed the passage of the eye wall over Naples just prior to the landfall of Hurricane Wilma. As the hurricane continued its northeastward track across southern Florida, a bow echo developed within the southern portion of the eye wall over western Dade County between 1130 and 1200 UTC. By 1150 UTC, as portrayed in Figure 6, bookend vortices became apparent on the northern and southern ends of the bow echo.

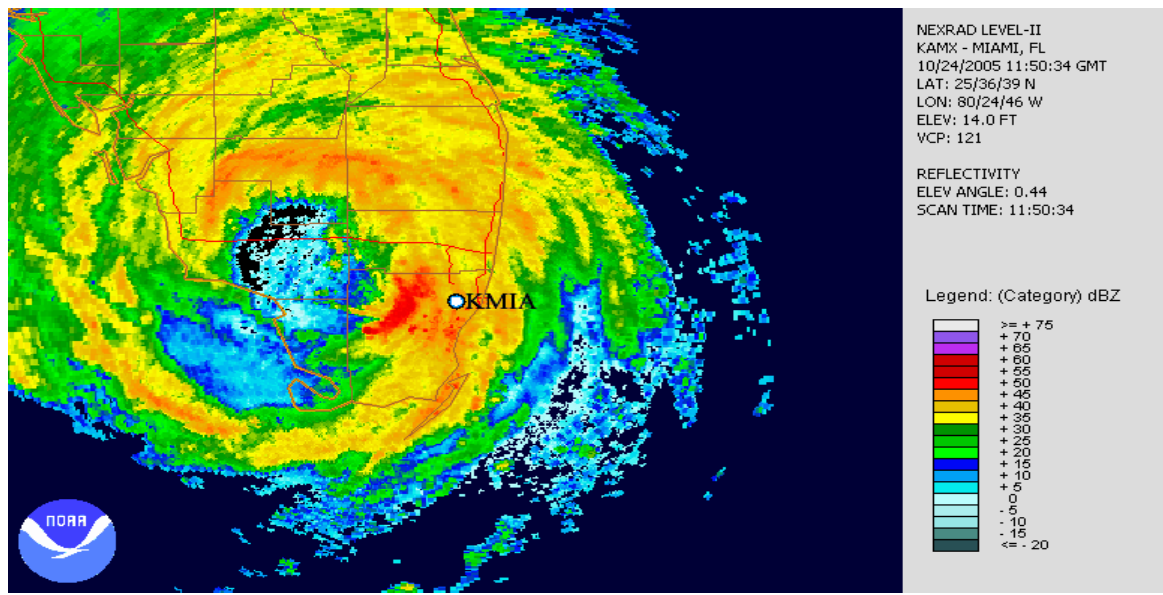

Figure 6. Miami, Florida NEXRAD reflectivity imagery at 1150 UTC 24 October 2005.

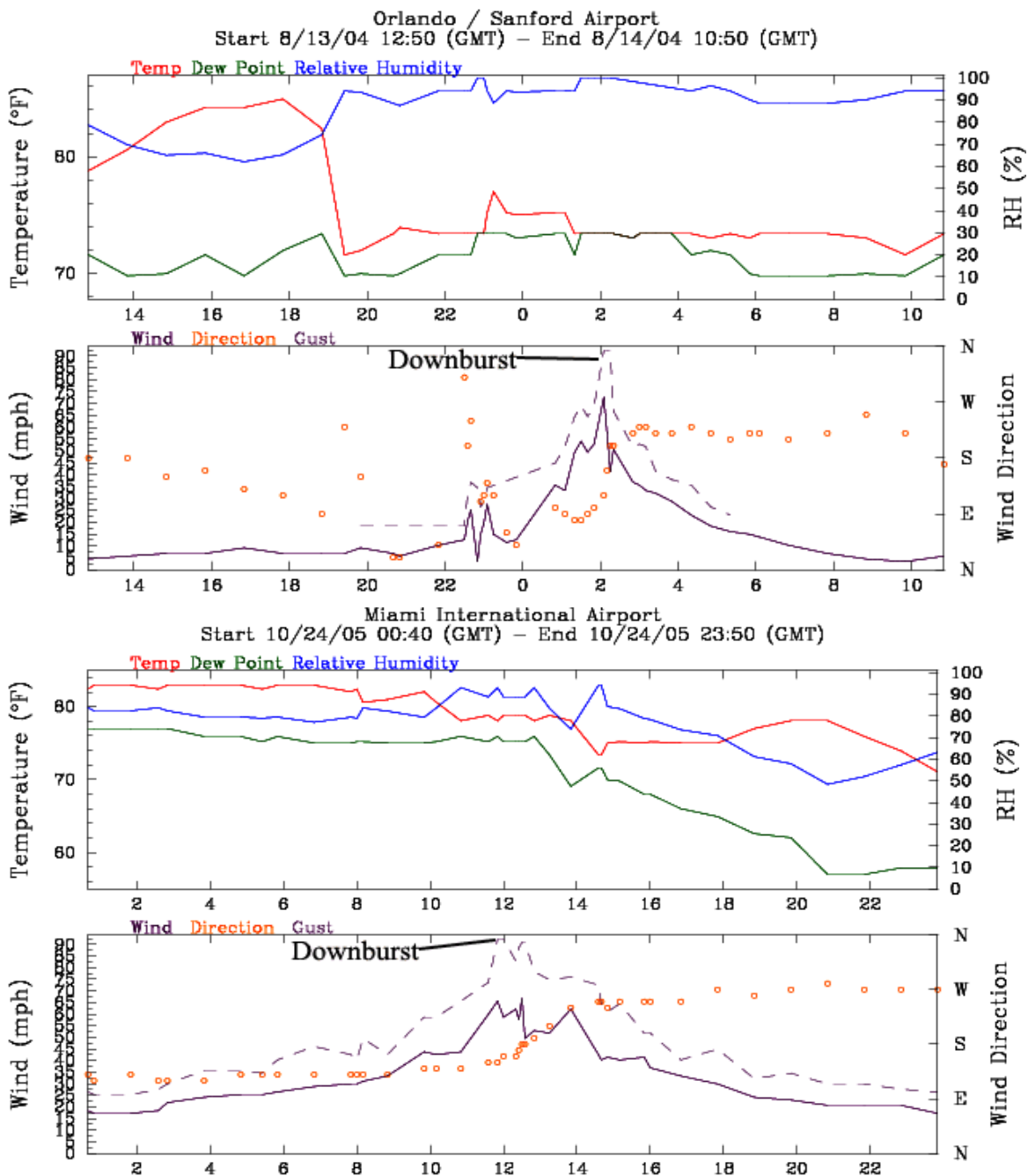

Figure 7.  Meteograms from Sanford (top) and Miami International Airport (bottom).

Shortly thereafter, a severe downburst was observed at Miami International Airport with a wind gust of 80 knots as the northern bookend vortex passed in close proximity to the airport. An interesting feature also evident in the radar image was the appearance of a spearhead echo embedded within the bookend vortex at the time of downburst observation.  In a similar manner to the downburst observed at Sanford in association with Hurricane Charley, Figure 7 displayed a sharp peak in wind speed that marked the occurrence of the downburst.  During the previous afternoon, as indicated in Figure 5, moderate WMSI values (>50) were indicated over the Miami area.

**4. Summary and Conclusions**

Hurricanes Charley and Wilma were notable in their generation of severe downbursts. Also interesting was the development of bow echoes within the remnant eye walls of both tropical cyclones. Observations of landfalling tropical cyclones have identified that a significant increase in low-level vertical wind shear occurs as a hurricane moves over land (Burpee 1986). Weisman (1993), in his study of severe, long-lived bow echoes, noted that environmental conditions required for the development of severe bow echoes include static instability (large CAPE) and the presence of strong vertical wind shear. The ambient environments along the tracks of these tropical cyclones after landfall were characterized by high WMSI values, resulting from the combination of large CAPE and the presence of a mid-tropospheric low theta-e layer, and vertical wind shear. Large CAPE was instrumental in the development intense updrafts within convective bands. In addition, in a manner similar to that described by Weisman (1993), intense vertical wind shear promoted the development of strong rear-inflow jets as well as bookend vortices. The bookend vortices associated with each bow echo served to enhance the strength of the rear-inflow jets, and hence, the channeling of low theta-e air into the mid-levels of the convective bands. The entrainment of relatively dry air (especially present within the eye) resulted in evaporative cooling and the generation of large negative buoyancy, and as a consequence, generating intense downbursts. Downward transport of high-momentum air into the boundary layer (Burpee 1986) may have also been a factor in the generation of strong winds associated with the pre-hurricane squall line of Charley as well as within the eye walls of both Hurricanes Charley and Wilma. Weisman (1993) also noted that a bookend vortex is also a favored location for downburst generation. Assuming that a change in air mass characteristics did not take place prior to hurricane landfall, the GOES WMSI product demonstrated effectiveness in indicating conditions favorable for downburst generation within tropical cyclones. Investigation of future landfalling tropical cyclones should serve to provide further support for this initial finding.

## 5. References

Atkins, N.T., and R.M. Wakimoto, 1991: Wet microburst activity over the southeastern United States: Implications for forecasting. *Wea. Forecasting*, **6**, 470-482.

Burpee, R.W., 1986: Mesoscale Structure of Hurricanes. In Mesoscale Meteorology and Forecasting. P.S. Ray (Ed.), American Meteorological Society, Boston, 311-330.

Pryor, K.L., and G.P. Ellrod, 2004: WMSI - A New Index For Forecasting Wet Microburst Severity. *National Weather Association Electronic Journal of Operational Meteorology*, 2004-EJ3.

Przybylinski, R.W., 1995: The bow echo. Observations, numerical simulations, and severe weather detection methods. *Wea. Forecasting*, **10**, 203-218.

Weisman, M.L., 1993: The genesis of severe, long-lived bow echoes. *J. Atmos. Sci.*, **50**, 645-670.